# Uncovering Social Network Sybils in the Wild

Zhi Yang[§], Christo Wilson[†], Xiao Wang[§‡], Tingting Gao[‡], Ben Y. Zhao[†], and Yafei Dai[§]
[§]Computer Science Department, Peking University, Beijing, China
[†]Computer Science Department, U. C. Santa Barbara, CA, USA
[‡]Renren Inc.
{*yangzhi, wangxiao, dyf*}@net.pku.edu.cn, {*bowlin, ravenben*}@cs.ucsb.edu, tingting.gao@renren-inc.com

## ABSTRACT

Sybil accounts are fake identities created to unfairly increase the power or resources of a single malicious user. Researchers have long known about the existence of Sybil accounts in online communities such as file-sharing systems, but have not been able to perform large scale measurements to detect them or measure their activities. In this paper, we describe our efforts to detect, characterize and understand Sybil account activity in the Renren online social network (OSN). We use ground truth provided by Renren Inc. to build measurement based Sybil account detectors, and deploy them on Renren to detect over 100,000 Sybil accounts. We study these Sybil accounts, as well as an additional 560,000 Sybil accounts caught by Renren, and analyze their link creation behavior. Most interestingly, we find that contrary to prior conjecture, Sybil accounts in OSNs do not form tight-knit communities. Instead, they integrate into the social graph just like normal users. Using link creation timestamps, we verify that the large majority of links between Sybil accounts are created accidentally, unbeknownst to the attacker. Overall, only a very small portion of Sybil accounts are connected to other Sybils with social links. Our study shows that existing Sybil defenses are unlikely to succeed in today's OSNs, and we must design new techniques to effectively detect and defend against Sybil attacks.

## 1. INTRODUCTION

Sybil attacks [4] are one of the most prevalent and practical attacks against distributed systems. In this attack, a malicious user creates multiple fake identities, known as Sybils, to unfairly increase their power and influence within a target community. Distributed systems are ill-equipped to defend against this attack, since determining a tight mapping between real users and online identities is an open problem. To date, researchers have demonstrated the efficacy of Sybil attacks against P2P systems [9], anonymous communication networks [1], and sensor networks [12].

Recently, online social networks (OSNs) have also come under attack from Sybils. Researchers have observed Sybils forwarding spam and malware on Facebook [5] and Twitter [6], as well as infiltrating social games [11]. Looking forward, Sybil attacks on OSNs are poised to become increasingly widespread and dangerous as more people come to rely on OSNs for basic online communication [8, 10] and as replacements of news outlets [7].

To address the problem of Sybils on OSNs, researchers have developed algorithms such as SybilGuard [22], SybilLimit [21], SybilInfer [3], and SumUp [15] to perform decentralized detection of Sybils on social graphs. These systems detect Sybils by identifying tightly connected communities of Sybil nodes [16]. However, to date no large scale studies have been performed to characterize the behavior of Sybils on OSNs in the wild. Thus, the assumptions underlying these algorithms remain untested.

In this paper, we describe our efforts to detect, characterize and understand Sybil account activity in Renren, the largest OSN in China. In Section 2, we use ground truth data on Sybils provided by Renren Inc. to characterize Sybil behavior. We identify several behavioral attributes that are unique to Sybils, and leverage them to build a measurement based, real-time Sybil detector. Our detector is currently deployed on Renren's production systems, and between August 2010 and February 2011 it led to the identification and banning of over 100,000 Sybil accounts.

In Section 3 we analyze the graph structural properties of Sybils on Renren, based on the 100,000 Sybils identified by our detector, as well as 560,000 more identified by Renren using prior techniques. Most interestingly, we find that contrary to prior conjecture, Sybil accounts in Renren do not form tight-knit communities: >70% of Sybils do not have *any* social edges to other Sybils at all. Instead, attackers use snowball sampling techniques to identify and send friend requests to popular users, since these users are more likely to accept requests from strangers. This strategy allows Sybil accounts to integrate seamlessly into the social graph.

We analyze the remaining 30% of Sybils that are friends with other Sybils, and discover that 69% (65,000 accounts) form a single connected component. By analyzing the creation timestamps of these edges, we determine that this component formed accidentally, and not due to coordinated efforts by attackers. We manually analyze several popular Sybil management tools, and show that large Sybil components form naturally due to bias in the snowball sampling techniques these tools use to locate targets for friending.

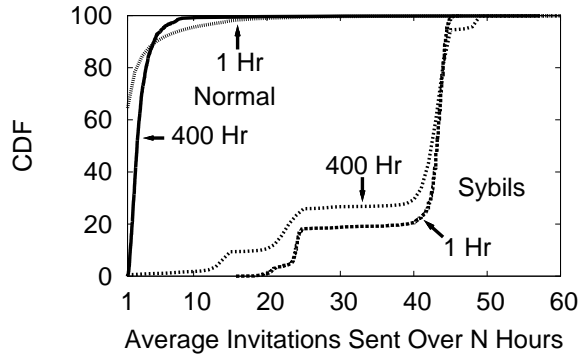

**Figure 1: Average friend invitation frequency over two time scales.**

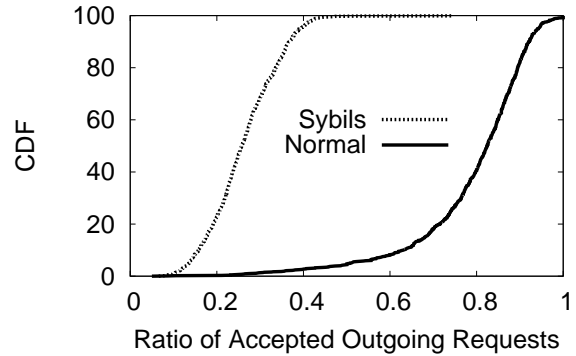

**Figure 2: Ratio of accepted *outgoing* friend requests.**

Our analysis of Sybil behavior and characteristics demonstrates that existing Sybil defenses are unlikely to succeed on today's OSNs. This opens the door for the development of new techniques to effectively detect and defend against Sybil attacks.

## 2. DETECTING SYBILS

In this section, we set the backdrop for our data analysis. First, we briefly introduce the Renren online social network and describe the role of Sybil accounts in Renren. Second, we describe experiments characterizing Sybil accounts on a verified ground-truth dataset provided by Renren. Finally, we describe and build a real-time Sybil account detector deployed on Renren, and show how it led to the large Sybil dataset we analyze in the remainder of the paper.

### 2.1 The Renren Network and Sybil Accounts

With 120 million users, Renren[1] is the largest and oldest online social network in China, and provides functionality and features similar to Facebook. Like Facebook, Renren first started in 2005 as a social network for college students in China, then saw its user population grow exponentially once it opened its doors to the non-student population. Like Facebook, Renren users maintain personal profiles, upload photos, write diary entries (blogs), and establish bidirectional social links with friends. The most popular type of user activity is sharing blog entries, which can be forwarded across multiple social hops much like "retweets" on Twitter.

As its user population has grown, Renren has become an attractive venue for companies to disseminate information about their products and activities. Users can become friends with pages that represent commercial companies such as Disney and McDonalds, from which users receive real-time news and updates of events on their home page. This has created opportunities for Sybil accounts to spam advertisements for companies, a growing trend observed by the analytics team at Renren. The increased prevalence of spam on Renren mirrors similar findings from Facebook [5] and Twitter [6].

[1] http://www.renren.com

To effectively attract friends and disseminate advertisements, most Sybil accounts on Renren blend in extremely well with normal users. They tend to have completely filled user profiles with realistic background information, coupled with attractive profile photos of young women or men, making their detection quite challenging.

Before this project, Renren had already deployed a suite of orthogonal techniques to detect Sybil accounts. To improve security for their users, Renren began a collaborative project with our research team in December 2010 to augment their detection systems with a systematic, real-time solution. To support the project, Renren provided full access to user data and operational logs on their servers, as well as allowing us to test and deploy research prototypes of Sybil detectors on their operational network.

### 2.2 Characterizing Sybil Accounts

Our approach to building a real-time Sybil detector begins by first identifying features that distinguish Sybil accounts from normal users. To help, Renren provided us with two sets of user accounts, containing 1000 Sybil accounts and 1000 non-Sybil accounts, respectively. The Sybil accounts were previously identified using existing mechanisms. A volunteer team carefully scrutinized all accounts in both sets to confirm they were correctly classified by looking over detailed profile data, including uploaded photos, messages sent and received, email addresses, and shared content (blogs and web links).

Using this dataset as our ground truth, we searched for behavioral attributes that may serve to identify Sybil accounts. After examining a wide range of attributes, we found four potential identifiers. We describe them each in turn, and illustrate how they characterize accounts in our ground truth dataset.

**Invitation Frequency.** Invitation frequency is the number of friend requests a user has sent within a fixed time period (*e.g.*, an hour). Figure 1 shows the friend invitation frequency of our dataset, averaged over long term (400 hour) and short term (1 hour) time scales. Since adding friends is a goal for all Sybil accounts, they are much more aggres-sive in sending requests than normal users. There is a clear separation: accounts sending more than 20 invites per time interval are Sybils. This result holds true at both long and short time scales, meaning that invitation frequency can be used to detect Sybil behavior without significant delays. For example, a threshold of 40 requests/hour can identify ≈70% of Sybils with no false positives.

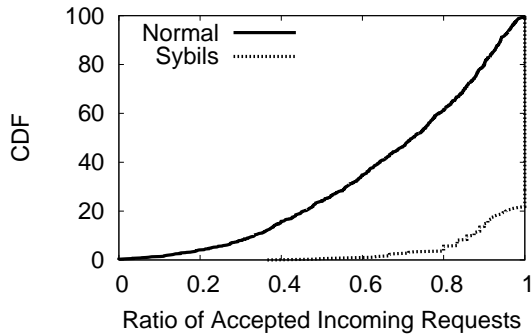

Figure 3: Ratio of accepted *incoming* friend requests.

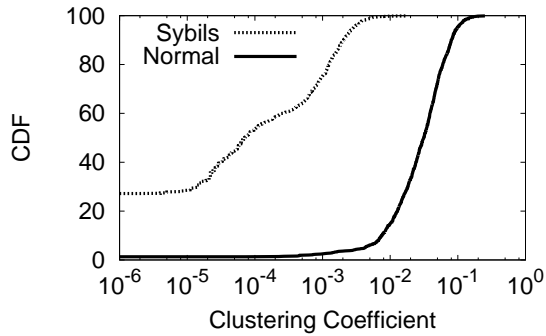

Figure 4: CDF of clustering coefficient for users' 50 first friends.

**Outgoing Requests Accepted.** A second distinguishing feature is the fraction of outgoing friend requests confirmed by the recipient. The CDF shown in Figure 2 shows a distinct difference between Sybils and normal users. In general, non-Sybil users generally have high accepted ratios with an average of 79%. On average, however, only 26% of all friend requests sent by Sybil accounts are accepted. This is unsurprising, since normal users typically send invites to people with whom they have prior relationships, whereas Sybils target strangers.

Despite prior studies that show users accept requests indiscriminately [13, 14], our results show that most users can still effectively identify and decline invitations from Sybil accounts. The fact that some users still accept requests from Sybil accounts is explained by two factors. First, most Sybils target members of the opposite sex by using photos of attractive young men and women in their profiles. While women make up 46.5% of the overall Renren user population, they make up 77.3% of the 1000 Sybil accounts in our dataset. Second, Sybils typically target popular users with numerous friends who are more likely to be open or careless about accepting friend requests from strangers. We further explore this point in Section 3.4.

**Incoming Requests Accepted.** Figure 3 plots a CDF of users by the fraction of incoming friend requests they accept. The incoming requests accepted by non-Sybil users are spread across the board. In contrast, Sybil accounts are nearly uniform in that they accept all incoming friend requests, *e.g.* 80% of Sybils accepted all friend requests. In fact, many of the Sybils with <100% accept rate fall into this category because Renren banned them before they could respond to all outstanding requests. However, since Sybil accounts receive few friend requests, this mechanism can incur a significant delay before detecting Sybils.

|  |  | SVM Predicted | | Threshold Predicted | |
|---|---|---|---|---|---|
|  |  | Sybil | Non-Sybil | Sybil | Non-Sybil |
| **True** | Sybil | 98.99% | 1.01% | 98.68% | 1.32% |
|  | Non-Sybil | 0.66% | 99.34% | 0.5% | 99.5% |

Table 1: Performance of SVM and threshold classifiers.

**Clustering Coefficient.** The clustering coefficient (cc) is a common graph metric that measures the mutual connectivity of a user's friends. Since normal users tend to have a small number of well-connected social cliques, we expect them to have much higher cc values than Sybil accounts, which are likely to befriend users with no mutual friendships. Figure 4 plots the CDF of cc values for each user's first 50 friends (sorted by time). As expected, non-Sybil users have cc values orders of magnitude larger than Sybil users (average cc values of 0.0386 and 0.0006 respectively). Since cc can be computed based on invitations only (*i.e.* user responses are not required) it can potentially perform well as a real-time Sybil detection metric.

### 2.3 Building and Running a Sybil Detector

Our analysis results seem to indicate that a threshold based scheme can effectively detect most Sybil accounts. Our next step is to verify this assertion by comparing the efficacy of a simple threshold detection approach against a more complex learning algorithm, *i.e.* a support vector machine.

We apply a support vector machine (SVM) classifier to our ground truth dataset of 1000 normal users and 1000 Sybils. We randomly partition the original sample into 5 sub-samples, 4 of which are used for training the classifier, and the last used to test the classifier. The results in Table 1 show that the classifier is very accurate, correctly identifying 99% of both Sybil and non-Sybil accounts. We compare these results to those of a threshold-based detector: *outgoing requests accepted ratio* $< 0.5 \wedge$ *frequency* $< 20 \wedge cc < 0.01$. Our results show that a properly tuned threshold-based detector can achieve performance similar to the computationally expensive SVM.

**Real-time Sybil Detection.** Our analytical results us-

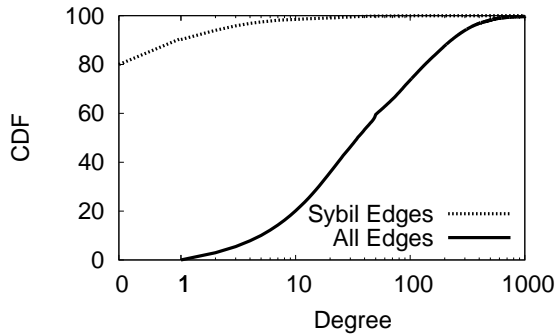

**Figure 5: The degree of Sybil accounts.**

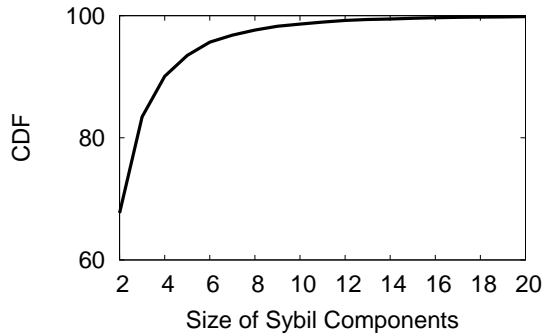

**Figure 6: The size of connected Sybil components.**

ing the ground-truth dataset led to the design of an adaptive, threshold-based Sybil detector that identifies Sybil accounts in near real-time. The detector monitors all accounts using a combination of friend-request frequency, outgoing request acceptance rates, and clustering coefficient. It uses an adaptive feedback scheme to dynamically tune threshold parameters on the fly[2].

After offline testing, Renren deployed our Sybil detection mechanism in late August 2010, and it has been in continuous operation ever since. From August 2010 to February 2011, Renren administrators used our mechanism to detect and subsequently ban ∼100,000 Sybil accounts in Renren. In addition to these accounts, Renren provided us with data on ∼560,000 accounts that were detected and banned using prior techniques from 2008 to February 2011. For the remainder of this paper, we will use all of these Sybil accounts (660,000 in all) to study the behavior of Sybil accounts.

## 3. SYBIL TOPOLOGY

In this section we analyze the graph topological characteristics of Sybil accounts on Renren. In particular, we are interested in analyzing whether Sybils in the wild are vulnerable to identification using the community-based Sybil detectors that have been proposed by researchers.

We begin the section with an overview of community-based Sybil detectors. We describe the algorithms they use to detect Sybils, and the key assumptions they make about Sybil behavior that enable them to function. Next, we analyze the degree distribution of Sybil accounts and demonstrate that, contrary to expectations, the vast majority do not form social links with other Sybil accounts. Next, we analyze connected components of Sybils. Temporal analysis of social links between Sybils indicates that these components formed randomly by accident, rather than intentionally due to the actions of an attacker. Finally, we examine popular tools used to create Sybils on Renren in order to explain how Sybil components naturally form.

### 3.1 Sybil Community Detectors

---
[2]We omit details of the adaptive scheme for Renren's security and confidentiality.

SybilGuard [22], SybilLimit [21], SybilInfer [3], and SumUp [15] are all algorithms for performing decentralized detection of Sybil nodes on social graphs. At their core, all of these algorithms are based on two assumptions of Sybil and normal user behavior:

1. Attackers can create unlimited Sybils and form edges between them. Edges between Sybils are beneficial since they make Sybils appear more legitimate to normal users.
2. The number of edges between Sybils and normal users will be limited, since normal users are unlikely to accept friend requests from unknown strangers.

Under these assumptions, Sybils tend to form tight knit clusters, since the number of edges between Sybils is greater than the number of edges connecting to normal users. We refer to edges between Sybils as *Sybil edges*, while edges connecting Sybils and normal users are called *attack edges*.

Sybil detection algorithms identify Sybil clusters by locating the small number of edge cuts that separate the Sybil region from the social graph. SybilGuard, SybilLimit, and SybilInfer all leverage specially engineered random walks for this purpose, while SumUp uses a max-flow approach. Although all of these algorithms are implemented differently, it has been shown that they all generalize to the problem of detecting communities of Sybil nodes [16].

Although these four algorithms have been shown to work on synthetic graphs (*i.e.* real social graphs with Sybil communities artificially injected), to date no studies have demonstrated their efficacy at detecting Sybils in the wild. In the following sections, we examine the characteristics of Sybils on Renren in order to ascertain whether they are amenable to identification by community-based Sybil detectors.

### 3.2 Sybil Edges

We begin our analysis of Sybil topology by examining the degree distribution of Sybil accounts on Renren. Our goal is to test the most basic assumption of community-based Sybil detectors: do Sybils in the wild form tight-knit communities? In order for Sybils to cluster, they must have at least one edge connecting to another Sybil, otherwise they will be disconnected.

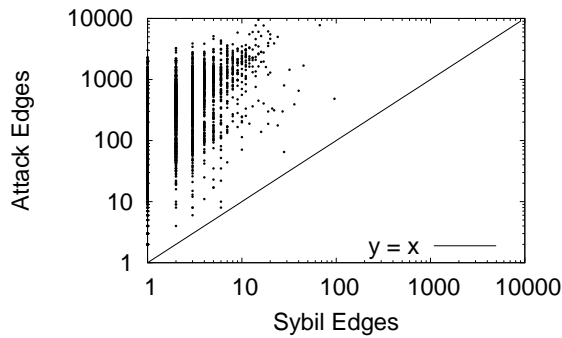 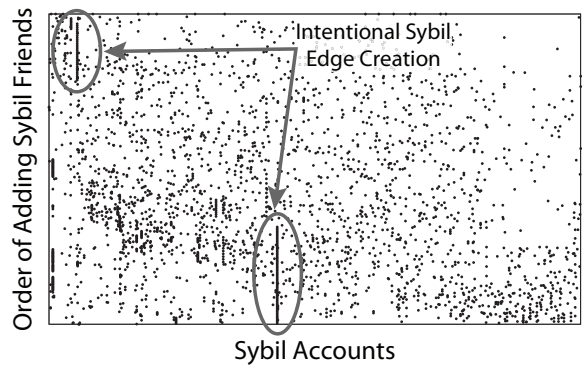

Figure 7: Scatter plot of Sybil edges vs. attack edges for Sybil components on Renren.

Figure 8: The order of adding Sybil friends for 1,000 Sybils. Each column represents an individual Sybil.

| Sybils | Sybil Edges | Attack Edges | Audience |
|---|---|---|---|
| 63,541 | 134,941 | 9,848,881 | 6,497,179 |
| 631 | 1153 | 1,040,745 | 21,014 |
| 68 | 67 | 7,761 | 7,702 |
| 51 | 50 | 15,349 | 15,179 |
| 37 | 40 | 14,431 | 13,886 |

Table 2: Statistics for the five largest Sybil components.

Figure 5 shows the degree distribution of all 667,723 Sybil accounts. When all edges are considered, the degree distribution is unremarkable: it follows the same general trend that has been observed on numerous other OSNs [19].

However, when we restrict the distribution to only plot edges between Sybils, we discover an unexpected result: only 20% of Sybils are friends with one or more other Sybils. This indicates that the vast majority of Sybil nodes do not demonstrate any sort of clustering behavior with other Sybils. Rather, most Sybils form only attack edges, and thus totally integrate into the normal social graph.

### 3.3 Sybil Communities

We now shift our focus to the minority of Sybils that do connect to other Sybils. Although we can conclude from Figure 5 that most Sybils in the wild do not obey the key assumption of community-based Sybil detectors, it is still possible that the connected minority are vulnerable to community detection. Thus, we now seek to answer the following questions: what are the characteristics of Sybil communities on Renren, and would community-based Sybil detectors be able to identify them?

To bootstrap our analysis, we construct a graph consisting solely of Sybils with at least one edge to another Sybil. The resulting graph is highly fragmented: it consists of 7,094 separate connected components. Figure 6 shows the size distribution of these Sybil components. As expected, the distribution is heavy tailed: 98% of Sybil components have less than 10 members. However the vast majority of Sybil accounts belong to a single, large connected component. Table 2 lists the details for the five largest Sybil components.

In order for Sybil communities to be identifiable by existing algorithms, they must form tight knit communities. Put another way, the number of Sybil edges inside the community must be greater than the number of attack edges that connect to the normal population. However, as shown in Table 2, this assumption does not hold for the largest Sybil components on Renren.

Figure 7 shows a scatter plot comparing the number of Sybil edges and attack edges in each Sybil component on Renren. All components are above the 45° line, meaning that they have more attack edges than Sybil edges. Thus, no components meet the requirements for detection using existing community-based Sybil identification algorithms.

### 3.4 Sybil Edge Formation

We now examine the processes driving the formation of Sybil edges on Renren. In particular, we seek to determine if edges between Sybil nodes are intentionally created by attackers. If they are, then this means that community detection may still be a viable approach to detecting Sybils on OSNs. However, if Sybil edges are not created intentionally, then this raises a new question: what process drives the accidental creation of Sybil edges?

**Temporal Characteristics.** One simple litmus test for identifying intentional Sybil edge creation is examining the order in which edges were established. If Sybil edges are formed intentionally by attackers, then we would expect to see them created sequentially, before friend requests are sent out to normal users.

Figure 8 shows the order in which edges were created for 1,000 random Sybils drawn from the largest Sybil component on Renren (containing 63,541 Sybils). For each Sybil $i$ with $n$ edges, we construct the sequence $\langle f_1, f_2, \ldots, f_n \rangle$, where $f_i$ is an edge, and the sequence is sorted chronologically by creation time. Each column of the figure shows the sequence of edge creations for a particular Sybil, with black dots representing Sybil edges.

As shown in Figure 8, the order of Sybil edge creation is almost uniformly random. For the most part, Sybil edges form randomly over the course of each Sybil's life. This in-dicates that the vast majority of Sybil edges in the large component were formed accidentally: attackers had no intention to link Sybils together and form a connected component. Intentional created connections between Sybils appear as solid vertical lines in the graph, which are observed for a handful of accounts in our sample. We highlight those accounts in Figure 8 by circling them.

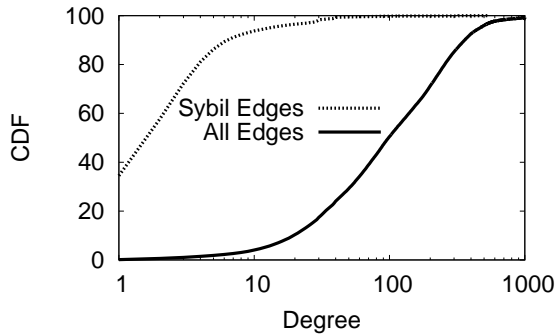

**Figure 9: Degree distribution of the largest Sybil component.**

| Tool Name & URL | Platform | Cost |
|---|---|---|
| Renren Marketing Assistant V1.0 http://www.duote.com/soft/30348.html | Windows | $37 |
| Renren Super Node Collector V1.0 http://www.snstools.com/snstool/86.html | Windows | Contact Author |
| Renren Almighty Assistant V5.8 http://www.sns78.com/ | Windows | Contact Author |

**Table 3: Popular Sybil creation and management tools.**

**Sybil Degree.** In order to reinforce the idea that the vast majority of Sybil edges in the large component are not intentionally created, we plot the degree distribution of the large component in Figure 9. 34.5% of Sybils only connect to 1 other Sybil, and 93.7% connect to ≤10. It is unlikely that an attacker would expend the effort to link their Sybils in such a loose way, since these low edge counts are not high enough to make Sybils appear legitimate to normal users.

**Snowball Sampling.** At this point we have established that attackers do not create the vast majority of Sybil edges intentionally; instead, they appear to occur randomly by accident. To understand how this happens, we conducted a survey of three popular software tools used to create and manage Sybil accounts on Renren. The details for each tool are given in Table 3. These tools advertise that they select targets for friending by performing snowball sampling on the social graph to locate popular users.

Based on the advertised functionality of these tools, we can surmise that Sybil edges are created accidentally due to two factors. First, the goal of Sybils is to accrue many friends by sending out numerous friend requests. If a Sybil is successful, it becomes popular by virtue of its large social degree. Second, the snowball sampling performed by Sybil management tools is intentionally biased towards locating popular users. Thus, it is likely that these tools will, unbeknownst to the attacker, occasionally select Sybil nodes to send friend requests to. As shown in Figure 3, Sybils almost always accept incoming friend requests, hence when this situation arises a Sybil edge is likely to be created.

## 4. RELATED WORK

**OSN Spam.** Recent studies have characterized the growing OSN spam problem on Facebook [5] and Twitter [6]. These studies rely on offline heuristics to identify spam content in status updates/tweets, as well as aberrant behavior that is indicative of spamming. The authors locate millions of spam messages on each OSN, and use them to analyze the large scale, coordinated spam campaigns. In contrast, our study is focused on the graph topological characteristics of malicious accounts, rather than spam content.

**OSN Spam Detection.** Various techniques borrowed from e-mail spam detection have been applied to OSN spam. Webb et. al. use honeypot accounts on MySpace to trap spammers who attempt to friend them [18]. Our results indicate that unless social honeypots are engineered to appear popular, they are unlikely to be targeted by spammers.

Other studies have leveraged Bayesian filters and SVMs to identify spammers on Twitter [2, 17, 20] and Facebook [14]. These techniques work well on Twitter, since Sybil friending behavior can be identified using publicly available following and followed information. However, detection on OSNs like Facebook and Renren is less successful, since the only publicly available indicators are laggy. Our Sybil detector overcomes this issue by leveraging friend invitation information that is only accessible from within Renren.

## 5. CONCLUSION

In this paper we make two contributions to the area of Sybil detection on OSNs. First, we use ground-truth data about the behavior of Sybils in the wild to create a measurement-based, real-time Sybil detector. We show that a computationally efficient, threshold-based classifier is sufficient to catch 99% of Sybils, with low false positive and negative rates. We have deployed our detector on Renren's production systems, and to date it has led to the identification and banning of over 100,000 Sybil accounts.

Our second contribution is a first-of-its-kind characterization of Sybil graph topology on a major OSN. Using edge creation information for 660,000 Sybil accounts on Renren, we show that Sybils in the wild do not obey behavioral assumptions that underlie previous work on decentralized Sybil detectors. Specifically, we demonstrate that the vast majority (80%) of Sybils do not connect to other Sybils. Even in cases where Sybils do form connected components, these clusters are loose, rather than tightly knit. Temporal analysis indicates that these Sybil edges are formed accidentally, rather than intentionally by attackers. These findings suggest that new approaches are needed to perform decentralized detection of Sybil accounts on OSNs.